\documentstyle[12pt]{article}

\begin{document}

\setcounter{page}{1}

\def\zm#1{#1_0}

\rightline{OSU--NT--94--07}
\vskip.5in

\centerline{\large {\bf Light-Cone Quantization of
Electrodynamics}\footnote{
Talk presented at ``Theory of Hadrons and Light-Front
QCD,'' Polana Zgorzelisko, Poland, August 1994.}
}
\vskip.3in
\centerline{ David G. Robertson }
\centerline{\it Department of Physics, The Ohio State University,
Columbus, OH 43210 }

\vskip.6 in
\centerline{\bf Abstract}
\vskip.1in
Light-cone quantization of (3+1)-dimensional electrodynamics is
discussed, using discretization as an infrared regulator and paying
careful attention to the interplay between gauge choice and boundary
conditions.  In the zero longitudinal momentum sector of the theory a
general gauge fixing is performed and the corresponding relations that
determine the constrained modes of the gauge field are obtained.  The
constraints are solved perturbatively and the structure of the theory
is studied to lowest nontrivial order.
\vskip.1in

\vskip.3in
{\bf 1. Introduction}
\vskip.1in

The primary motivation for considering light-cone quantization when
attempting to derive a constituent picture of hadrons from QCD is that
a single cutoff, removing states of small longitudinal momentum,
renders the vacuum of the theory completely trivial.  This immediately
results in a picture in which all partons in a physical hadronic state
are tied to the hadron, instead of being disconnected excitations in a
complicated vacuum.  The situation is quite different when quantizing
on a spacelike surface, where no simple cutoff can remove the states
that are kinematically allowed to mix with the bare Fock vacuum. In
this case a constituent description would necessarily be in terms of
quasiparticle states, i.e., collective excitations above a complicated
ground state.

One calculational scheme which accomplishes (in most cases) this
small-$k^+$ cutoff is ``discretized'' light-cone quantization (DLCQ).
In this approach one imposes periodicity conditions on the fields in
the $x^-$ direction, leading to a discrete set of allowed longitudinal
momenta.  For Fermi fields it is permissible to choose antiperiodic
boundary conditions so that there is no zero-momentum mode.  For
bosonic fields one must generally choose periodic boundary conditions,
but the zero mode is in most cases not an independent field; rather,
it is a constrained functional of the other, dynamical fields in the
theory.  The bare vacuum is thus the only state in the theory with
$k^+=0,$ and so is an exact eigenstate of the full Hamiltonian.  The
price we pay for this simplification is that we must solve the
constraints that determine the zero modes.

The exception to this is gauge theories.  Here one finds that the zero
modes of $A^i$ and $A^-$ are constrained fields, but certain of the
zero modes of $A^+$ can in fact be dynamical.  Furthermore, these
cannot be removed from the theory by a gauge choice.  Thus in this
case there are $k^+=0$ particle states that can in principle mix with
the bare vacuum to give a nontrivial physical vacuum state, although
there are many fewer such states than typically occur in equal-time
quantization.  In this case the nontrivial vacuum must either be
confronted or further {\it ad hoc} truncations must be made in order
to obtain a trivial vacuum.

In this talk I summarize recent work with A. Kalloniatis on the DLCQ
of Abelian gauge theory with fermions [1].  The central problem is
disentangling the constrained and dynamical zero modes in the context
of a particular gauge fixing of the theory.  After this has been done,
we then face the problem of solving the relations that determine the
constrained zero modes so that the Hamiltonian can be written down.  I
shall describe here the formulation of the theory to lowest order in
perturbation theory.

\vskip.3in
{\bf 2. Gauge Fixing and the Zero Momentum Modes }
\vskip.1in

In DLCQ the theory is defined in a light-cone box, with $-L_\perp\leq
x^i\leq L_\perp$ and $-L\leq x^-\leq L,$ and with some boundary
conditions imposed on the fields.  Because the gauge field couples to
a fermion bilinear, which is necessarily periodic in all coordinates,
$A_\mu$ must be taken to be periodic in both $x^-$ and $x_\perp$.  We
have more flexibility with the Fermi field, and it is convenient to
choose this to be periodic in $x_\perp$ and antiperiodic in $x^-.$
This eliminates the zero longitudinal momentum mode while still
allowing an expansion of the field in a complete set of basis
functions.

These functions are taken to be plane waves, and for periodic fields
there will of course be zero-momentum modes.  In previous work on the
formulation of QED in DLCQ [2] the zero modes were discarded, and only
the nonzero, or ``normal,'' modes were retained.  This part of the
theory is essentially unchanged.  The ``global'' zero modes---those
independent of all spatial coordinates---require special treatment,
and the reader is referred to Appendix B of ref. [1] for more details.
(It is shown there that they are irrelevant for the calculations we
shall present here.)  The quantities that will be of central interest
to us here are the ``proper'' zero modes, which are independent of
$x^-$ but {\it not} independent of $x_\perp.$ Specifically, for a
periodic quantity $f$, the proper zero mode is defined as
\begin{equation}
\zm{f}(x_\perp) \equiv \int_{-L}^L {dx^-\over2L} f(x^-,x_\perp)
- \int_{-L}^L {dx^-\over2L}
\int_{-L_\perp}^{L_\perp} {d^2x_\perp\over(2L_\perp)^2}
f(x^-,x_\perp)\; .
\end{equation}
Projecting the equation of motion $\partial_\mu F^{\mu\nu}=gJ^\nu$
onto the proper zero mode sector gives
\begin{equation}
-\partial_\perp^2\zm{A}^+ = g\zm{J}^+ \; ,
					\label{eq:zm2}
\end{equation}
\begin{equation}
-2(\partial_+)^2\zm{A}^+-\partial_\perp^2\zm{A}^-
-2\partial_i\partial_+\zm{A}^i = g\zm{J}^- \; ,
					\label{eq:zm3}
\end{equation}
\begin{equation}
-\partial_\perp^2\zm{A}^i+\partial_i\partial_+\zm{A}^+
+\partial_i\partial_j\zm{A}^j = g\zm{J}^i
\; .
					\label{eq:zm1}
\end{equation}
We first observe that eq. (\ref{eq:zm2}) is a constraint which
determines $\zm{A}^+$ in terms of $J^+$.  Eqs. (\ref{eq:zm3}) and
(\ref{eq:zm1}) then determine $\zm{A}^-$ and $\zm{A}^i$.  Thus all of
the proper zero modes are constrained fields.

Eq. (\ref{eq:zm2}) is clearly incompatible with the light-cone gauge
$A^+=0,$ which is most natural in light-cone analyses of gauge
theories.  Here we encounter a common problem in treating axial gauges
on compact spaces, which has nothing to do with light-cone
quantization {\it per se.} The point is that the $x^-$-independent
part of $A^+$ is in fact gauge invariant, since under a gauge
transformation
\begin{equation}
A^+\rightarrow A^++2\partial_-\omega\; ,
\end{equation}
where $\omega$ is a function periodic in all coordinates.  Thus it is
not possible to bring an arbitrary gauge field configuration to one
satisfying $A^+=0$ via a gauge transformation.  We can (and will) set
the normal mode part of $A^+$ to zero, which is equivalent to
\begin{equation}
\partial_-A^+=0\; .
					\label{eq:gauge1}
\end{equation}
This does not, however, completely fix the gauge---we are free to make
arbitrary $x^-$-independent gauge transformations without undoing eq.
(\ref{eq:gauge1}).  We may therefore impose further conditions on
$A_\mu$ in the zero mode sector of the theory.

To see what might be useful in this regard, let us consider solving
eq. (\ref{eq:zm1}) for the transverse zero modes.  By making use of
current conservation we can rewrite eq. (\ref{eq:zm1}) as
\begin{equation}
-\partial_\perp^2\Bigl(\delta^i_j-{\partial_i\partial_j\over\partial_\perp^2}
\Bigr)\zm{A}^j =
g\Bigl(\delta^i_j-{\partial_i\partial_j\over\partial_\perp^2}\Bigr)
\zm{J}^j\; ,
\end{equation}
the general solution of which is
\begin{equation}
\zm{A}^i = -g{1\over\partial_\perp^2}\zm{J}^i
+ \partial_i\varphi(x^+,x_\perp)\; .
\end{equation}
Here $\varphi$ must be independent of $x^-$ but is otherwise
arbitrary, and reflects the residual gauge invariance.  Imposing a
condition on, say, $\partial_i\zm{A}^i$ will uniquely determine this
function.

Now different choices for $\varphi$ merely correspond to different
gauge choices in the zero mode sector, so we expect that physical
quantities should be independent of the specific $\varphi$ we choose.
This is in fact the case.  It turns out, however, that the unique
choice which allows simple (free-field) commutation relations among
the fields is $\varphi=0.$ The easiest way to see this is to note that
for $\varphi\neq0$ the kinematical operators $P^i$ do not have their
free-field forms [1].  Thus it would be impossible to realize the
Heisenberg relation
\begin{equation}
[\psi_+,P^i]=-i\partial_i\psi_+
\end{equation}
with the usual anticommutation relation between $\psi_+$ and
$\psi_+^\dagger$.  The complicated anticommutator required could in
principle be determined, but this is not really necessary.  It is far
simpler to choose the gauge $\varphi=0$ and take the usual canonical
commutation relations among the fields, and this is what we shall do.

\vskip.3in
{\bf 3. Perturbative Formulation }
\vskip.1in

We shall now construct a perturbative solution of the constraints and
study the structure of the theory to lowest nontrivial order.  This
requires constructing the Hamiltonian through terms of ${\cal
O}(g^2),$ which in turn corresponds to taking the ${\cal O}(g)$
solutions for the zero modes $\zm{A}^i$ and $\zm{A}^+$.  We obtain
$\zm{A}^i$ at this order simply by setting $g=0$ in the current
$\zm{J}^i$.  $\zm{A}^+$ is given exactly by inverting the
$\partial_\perp^2$ in eq. (\ref{eq:zm2}).

With $\zm{A}^i$ and $\zm{A}^+$ in hand it is simple to construct
the Hamiltonian.  The contributions from the normal mode part of the
theory may be found in ref. [2].\footnote{
Note that the Hamiltonian given in ref. [2] should be symmetrized
under charge conjugation ($b\leftrightarrow d$).  This is the effect
of using the explicitly $C$-odd form of the current
$J^\mu={1\over2}[{\overline{\psi}},\gamma^\mu\psi]$, or equivalently
of properly symmetrizing products of noncommuting operators in the
construction of $P^-$.}
To this we must add the contributions from the proper zero modes,
which reduce to [1]
\begin{equation}
P^-_Z = {g^2\over2} \int dx^- d^2x_\bot
\Bigl({\partial_i\over{\partial_\bot^2}} \zm{J}^\mu \Bigr)
\Bigl({\partial_i\over{\partial_\bot^2}} {\zm{J}}_\mu \Bigr) \; ,
					\label{eq:zmcontri}
\end{equation}
where the currents are evaluated with $g=0.$ It is a straightforward
(if tedious) exercise to express $P^-_Z$ in the Fock representation.
We obtain various four-fermion operators, as well as fermion bilinears
(``self-induced inertias'') which arise when the four-point terms are
brought into normal order.

We can now compute with this Hamiltonian and study the effects of the
zero mode-induced interactions (\ref{eq:zmcontri}).  We shall here
describe the calculation of the fermion self-energy (eigenvalue of
$P^-$) at ${\cal O}(g^2)$.  This is not the only quantity to which
$P^-_Z$ contributes, of course.  The four-fermion operators in $P^-_Z$
will give rise to divergent contributions to the $e^+e^-\gamma$
vertex, and hence to the charge renormalization at this order.  They
will also contribute to tree-level scattering amplitudes, {\it etc.}

In fact, we shall consider only the part of the self-energy that is
quadratically divergent in a transverse momentum cutoff.  The
contributions coming from the normal mode sector of the theory may be
taken from the work of Tang, {\it et al.} [2] (subject to the caveat
in footnote 2).  It has the form
\begin{equation}
\delta P^- = c\Lambda_\perp^2\Biggl[
{2\ln2\over p^+} - {1\over (p^+)^2L}\Biggr]\; ,
					\label{eq:result}
\end{equation}
where $p^+$ is the momentum of the fermion and $c$ is a dimensionless
constant.

Two points should be made about this result.  The first is that it
represents a failure of chiral symmetry.  Corrections to the fermion
mass are supposed to be proportional to $m$, so that they vanish in
the chiral limit.  Here, however, there is a correction to $m$ [the
first term in eq. (\ref{eq:result})] that is completely independent of
$m$. Furthermore, the second term in eq. (\ref{eq:result}) does not
have the correct form to be interpreted as the redefinition of a
parameter in the Lagrangian.  It would have to be removed by a
noncovariant counterterm.

There is another quadratically divergent contribution to the
self-energy, however, coming from the zero mode-induced interactions
(\ref{eq:zmcontri}).  It reduces to
\begin{equation}
\delta P^-_{zm} = +c\Lambda_\perp^2\Biggl[{1\over (p^+)^2L}\Biggr]\; ,
\end{equation}
so that the noncovariant part of the quadratic divergence is in fact
canceled when the zero modes are retained.  Thus inclusion of the zero
modes renders the UV behavior of the theory more benign.  The
quadratic divergence proportional to ${1\over p^+}$ survives, unlike
in a continuum formulation (with a Lorentz-covariant regulator), but
this may be removed by a redefinition of the fermion kinetic mass.

\vskip.3in
{\bf 4. Discussion }
\vskip.1in

The box length $L$ is introduced to regulate the theory in the
infrared, and in principle it should be taken to infinity at the end
of the day.  More concretely, one should calculate physical quantities
of interest, which may be functions of the parameter $L$, and study
them in the limit $L\rightarrow\infty.$ For the energy of the
one-fermion state, for example, this means taking $L\rightarrow\infty$
with the $p^+$ of the state held fixed.  Thus the term in eq.
(\ref{eq:result}) that is canceled by the zero mode contribution is
explicitly $L$-dependent, in addition to being quadratically divergent
in $\Lambda_\perp$.  It is in some sense ``irrelevant''---it goes to
zero as the infrared cutoff is removed---but this is not the point we
wish to emphasize.  Rather it is that the zero mode-induced
interactions {\it removed} this $L$-dependence.  This suggests that at
least some of the zero mode interactions can be thought of as infrared
counterterms, that serve to remove dependence on the infrared cutoff
$L$.

In a theory like QED, in which we do not expect any vacuum structure
(in the sense of, e.g., vacuum expectation values or a
$\theta$-vacuum), we therefore expect that inclusion of the zero mode
interactions mainly helps to remove $L$-dependence from the solution
of the theory and so makes the infinite volume limit easier to reach.
There is numerical evidence that this is the case [3].  A useful
analogy might be with improved actions for lattice theories, in which
dependence on the lattice spacing $a$ is explicitly removed through
some order.  Calculations using the improved action can then be done
at a larger value of $a$ (fewer lattice points for a given physical
volume) for a fixed numerical accuracy.

It is clearly of interest to learn whether there are zero mode
interactions that are marginal or relevant in the renormalization
group sense.  This question can perhaps be addressed systematically
from the point of view of the light-cone power-counting analysis of
Wilson [4].

It is also important to develop more sophisticated, nonperturbative
methods for solving the constraints. For QED with a realistic value of
the electron charge, however, it is possible that a perturbative
treatment of the constraints could suffice; that is, that we could use
a perturbation theory to construct the Hamiltonian, and then
diagonalize it nonperturbatively.  This approach is similar in spirit
to that advocated in ref. [4], where the idea is to use a perturbative
realization of the renormalization group to construct an effective
Hamiltonian for QCD, which is then solved nonperturbatively.

We have not addressed here the general issue of removing dependence on
the transverse cutoff $\Lambda_\perp$ from the theory.  This is
perhaps the more difficult problem, and in fact it may make its
presence felt in calculations like those presented here.  The
constraint relations are strictly speaking ill-defined due to
transverse UV divergences, and how one regulates and removes these
presumably affects the treatment of the zero modes.  This may be an
important issue for theories with nontrivial UV divergences.

\vskip.3in
{\bf Acknowledgements}
\vskip.1in
It is a pleasure to thank the organizers of the conference for their
hospitality.  This work was supported by the National Science
Foundation under Grants Nos. PHY-9203145, PHY-9258270, and
PHY-9207889.

\vskip.3in
{\bf References}
\vskip.1in

1. A.C. Kalloniatis and D.G. Robertson, Heidelberg/OSU preprint
MPIH--V10--1994,\par\quad OSU--NT--94--03 ({\tt hep-th/9405176}), to
appear in Phys. Rev. D.

2. A.C. Tang, S.J. Brodsky, and H.-C. Pauli, Phys. Rev. D{\bf 44},
1842 (1991).

3. J.J. Wivoda and J.R. Hiller, Phys. Rev. D{\bf 47}, 4647 (1993).

4. K.G. Wilson, {\it et al.}, Phys. Rev. D{\bf 49}, 6720 (1994).

\end{document}